\documentclass[times, twoside]{zHenriquesLab-StyleBioRxiv}

\usepackage{amsmath,amsfonts,amssymb}
\usepackage{booktabs}
\usepackage{multirow}
\leadauthor{Jansen} 

\begin{document}

\title{Liver segmentation and metastases detection in MR images using convolutional neural networks}
\shorttitle{Liver segmentation and liver metastases detection}

\author[1,\Letter]{Mari\"elle J.A. Jansen}
\author[1,]{Hugo J. Kuijf}
\author[2,]{Maarten Niekel}
\author[2,]{Wouter B. Veldhuis}
\author[2,]{Frank J. Wessels}
\author[1,]{Max A. Viergever}
\author[1,]{Josien P.W. Pluim}

\affil[1]{UMC Utrecht and Utrecht University, Image Sciences Institute, P.O. Box 85500, Q02.445, 3508 GA Utrecht, the Netherlands}
\affil[2]{UMC Utrecht, Department of Radiology, P.O. Box 85500, Q02.445, 3508 GA Utrecht, the Netherlands}

\maketitle

\begin{abstract}
Primary tumors have a high likelihood of developing metastases in the liver and early detection of these metastases is crucial for patient outcome. We propose a method based on convolutional neural networks (CNN) to detect liver metastases. First, the liver was automatically segmented using the six phases of abdominal dynamic contrast enhanced (DCE) MR images. Next, DCE-MR and diffusion weighted (DW) MR images are used for metastases detection within the liver mask. The liver segmentations have a median Dice similarity coefficient of 0.95 compared with manual annotations. The metastases detection method has a sensitivity of 99.8\% with a median of 2 false positives per image. The combination of the two MR sequences in a dual pathway network is proven valuable for the detection of liver metastases. In conclusion, a high quality liver segmentation can be obtained in which we can successfully detect liver metastases.
\end {abstract}


\begin{corrauthor}
marielle\at isi.uu.nl
\end{corrauthor}

\section*{Introduction}
Primary tumors, such as neuroendocrine and colorectal tumors, have a high likelihood of developing metastases in the liver.\textsuperscript{1,2} Early detection of (new) liver metastases is crucial since it improves patient outcome.\textsuperscript{3-5} To follow disease progress, radiologists check for tumor growth and new (liver) metastases in computed tomography (CT) or magnetic resonance (MR) images.\par
While CT has long been the modality of choice in detecting and monitoring liver tumors, MRI has gained interest thanks to a better lesion-to-liver contrast and because it does not use ionizing radiation.\textsuperscript{6,7} Dynamic contrast-enhanced (DCE) MR images have a high sensitivity and specificity for visual detection of liver metastases. Moreover, the combination of DCE-MR and diffusion weighted (DW) MR images turned out to be even more effective in the visual detection of liver metastases and visual censoring of mimics.\textsuperscript{8,9,10} \par
The automatic detection and characterization of liver metastases remains a challenging task, given the heterogeneous appearance of liver metastases on MR images.\textsuperscript{11,12} Radiologists have a liver lesion detection rate between 87-95\%, when using both DCE-MR and DW-MR images.\textsuperscript{7,8,9} Automatic liver metastases detection could aid radiologists in finding metastases more efficiently and effectively.\par
The liver metastases detection method we propose is a two-step method: a liver segmentation step followed by a lesion detection step within the segmented liver. Most liver segmentation methods published have been developed for CT, with best results by convolutional neural network (CNN) based methods.\textsuperscript{13,14} Fewer methods have been developed for segmentation of the liver in MR images. Those known are primarily based on watershed\textsuperscript{15,16}, active contouring\textsuperscript{17}, atlases\textsuperscript{18}, or shape models\textsuperscript{19}.\par 
For automatic detection of liver lesions several methods have been proposed, all based on CT images.\textsuperscript{14, 20-23} MR data was employed by a few methods to detect hepatocellular carcinomas (either DCE-MRI\textsuperscript{24} or DW-MRI\textsuperscript{25}), but not for metastases.\par
Our aim is to develop and evaluate a two-step liver metastases detection method for MR images, based on fully convolutional neural networks (FCN). In the first step, a liver segmentation method utilizing the dynamic nature of the DCE-MR images is presented, based on previous work\textsuperscript{26}. This is followed by a method for the detection of metastases within the liver region using both DCE-MR and DW-MR images as input.

\section*{Data}
The study comprises MR data of 121 patients with a clinical focus on the liver from the University Medical Center Utrecht, the Netherlands, acquired between February 2015 and February 2018. The UMCU Medical Ethical Committee has reviewed this study and informed consent was waived due to its retrospective nature.\par
All patients underwent a clinical MR examination, including DCE-MR series and DW-MRI. The DCE-MR series was acquired in six breath holds with one to five 3D images per breath hold, with the following parameters: TE: 2.143 ms; TR: 4.524 ms; flip angle: 10 degrees. After acquiring the first image, gadobutrol (0.1 ml/kg Gadovist of 1.0 mmol/ml at 1 ml/s) was administered at once, followed by 25 ml saline solution at 1 ml/s. In total, 16 3D images per patient were acquired with 100 slices and matrix sizes of 256 $\times$ 256. Voxel size was 1.543 mm $\times$ 1.543 mm $\times$ 2 mm. \par
The DW-MR images were acquired with three b-values: 10, 150, and 1000 s/mm2, using a protocol with the following parameters: TE: 70 ms; TR: 1.660 ms; flip angle: 90 degrees. Each b-value image was acquired with 42 slices and matrix sizes of 256 $\times$ 256. Voxel size was 1.758 mm $\times$ 1.758 mm $\times$ 5 mm.

\subsection*{Pre-processing}
\subsubsection*{DCE-MR}
All DCE-MR data sets were corrected for motion using a groupwise registration.\textsuperscript{27} The groupwise registration method registers all images simultaneously to a common space by minimizing a cost function based on principle component analysis and applying a B-spline transformation. The registration is applied on four resolutions with 500 iterations each.  After registration, a zero-mean-unit-variance rescaling was applied to all intensity values between the 0\textsuperscript{th} and 99.8\textsuperscript{th} percentile of the intensity histogram of DCE-MR series. The 99.8\textsuperscript{th} percentile intensity was assumed to correspond to the contrast agent peak in the aorta.\par
The images were combined per phase by averaging the fourth dimension. The series started with the pre-contrast image, followed by the other phases: the early arterial phase, the late arterial phase, the hepatic/portal-venous phase, the late portal-venous/equilibrium phase and the late equilibrium phase. The six phases were used as input images for the FCN.

\subsubsection*{DW-MR}
The DW-MR images were nonlinearly registered to the DCE-MR space using elastix, a toolbox for linear and non-linear registration of medical images.\textsuperscript{28} The fixed image was the mean DCE-MR, obtained by averaging over the six phases. First a rigid transformation was applied on two resolutions with 2000 iterations each, followed by a b-spline transformation on one resolution with 1000 iterations and a grid spacing of 60$\times$60$\times$40 mm. Normalized mutual information was used as metric. A mask was used to focus the registration on the liver. The mask was obtained from the automatic liver segmentation, morphologically dilated with a 10$\times$10$\times$10 structuring element. A substantial dilation was chosen to make sure the boundary of the liver is included.\par
Ten out of the 121 MRI examinations were excluded from the study because of a failed registration of the DW-MRI on the DCE-MR series. The exclusion was based on visual evaluation.\par
The intensities of the registered DW-MR data set were also normalized with a zero-mean-unit-variance rescaling between the 0\textsuperscript{th} and 99.8\textsuperscript{th} percentile of the intensity histogram of the DW-MRI.

\subsection*{Annotations}
The liver was annotated in 55 DCE-MR series, of which only 16 series contained metastases that were segmented. Additionally, we included another 56 DCE-MR series in which the metastases were segmented, resulting in a total of 72 DCE-MR series with annotated metastases.

\subsubsection*{Liver segmentation}
Fifty-five dynamic contrast enhanced MR series were used for liver segmentation. The data sets were randomly divided in thirty-three data sets for training, three for validation, and nineteen data sets for testing. The validation set was used for tuning the hyperparameters and the evaluation of the CNN model during method development. The test set was used to evaluate the final CNN model in an independent manner.\par
The liver was manually contoured by two observers, see Fig. 3 for segmentation examples. The first observer annotated the training, validation, and test sets. The annotation of the test set is indicated as O1.1. The first observer repeated the annotations of the test set at least one week later (O1.2). The second observer annotated the test set once (O2). This was done to estimate the inter- and intra-observer agreement. A radiologist with more than ten years of experience in liver MR analysis verified all manual annotations and provided corrections where needed. Liver lesions were included in the annotation and this network was therefore trained to recognize liver lesions as liver tissue. The first set of annotations of the first observer was used as reference in the experiments.

\subsubsection*{Liver metastases detection}
Seventy-two MR data sets were used for the liver metastases detection. The data set included mainly colorectal metastases, neuroendocrine metastases, and some other metastasis types (i.e. other gastrointestinal metastases and breast metastases). The data sets were randomly divided in fifty-five data sets for training (n=50) and validation (n=5), and seventeen data sets for testing. The training and validation sets contained 334 metastases in total, with on average 6 metastases per liver (range: 1 - 31 metastases). The test set contained 86 metastases in total, with on average 5 metastases per liver (range: 1 - 32 metastases).\par
The metastases were manually annotated on the DCE-MR images by a radiologist in training and were verified by a radiologist with more than ten years of experience.

\section*{Methods}
\subsection*{Liver segmentation}
An FCN\textsuperscript{29} with dilated convolutions was implemented, for which the six phases of the DCE-MR images were the channels of the input image. The dilated FCN consisted of 9 convolutional layers in total. The first 7 layers had a 3$\times$3 convolution and 32 kernels. The dilation rates ranged from 1 to 16. The final two layers had a 1$\times$1 convolution. ReLU activation and batch normalization were used in all the convolutional layers, except for the final layer which had a softmax activation. A dropout layer was applied between the 8\textsuperscript{th} and the 9\textsuperscript{th} layer with a dropout rate of 0.5. This network had a receptive field of 67$\times$67 pixels. Figure 1 gives an overview of the network architecture. In our previous work on liver segmentation\textsuperscript{26}, we also considered the popular U-net architecture. However, this architecture had a slightly worse performance than the herein used FCN architecture.\par
The loss was calculated by a similarity metric based on the Dice similarity coefficient: $(2*X\cap Y+s)/(X^2+Y^2+s)$, where $X$ is the predicted segmentation, $Y$ is the ground truth mask, and $s$ is a small number to prevent dividing by zero (set to 1e-5).30 Glorot uniform\textsuperscript{31} was used as initializer and Adam as optimizer with a learning rate of 0.001. The network was trained for 100,000 iterations, with six images per mini batch. The total number of 2D slices for training was 3300 slices. No data augmentation was applied. \par
The data was processed by the network per 2D slice consisting of 6 channels. For the evaluation of the liver segmentation, the probability output was post-processed to a binary image. The threshold of 0.5 was applied to the probability output of the network. After that 3D hole filling was performed, so that all holes caused by liver lesions were filled. To remove small spurious segmentations, the largest connected component was selected as the final segmentation.

\begin{figure}
	\begin{center}
		\begin{tabular}{c} 
			\includegraphics[width=8.5cm]{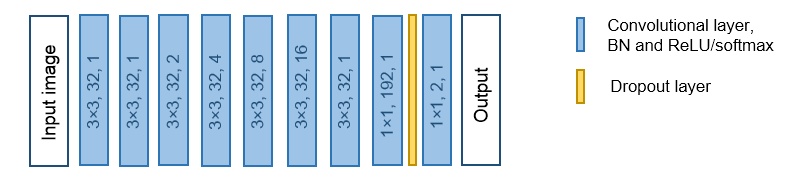}
		\end{tabular}
	\end{center}
	\caption[example]
	{ \label{fig:Fig1} 
		 Network architecture for liver segmentation. The blue blocks represent the convolutional layers, batch normalization (BN) and ReLU or softmax activation. The size of the kernel of each convolutional layer is given in the block, followed by the number of kernels and the dilation rate of the kernel.}
		 
\end{figure}

\subsection*{Liver metastases detection}
A dual pathway FCN was implemented, for which the input images of one path were the six DCE-MR phase images and for the other path the three DW-MR images. The six (or three) 2D images were combined to one input image, with six (or three) channels. Each pathway had 13 convolutional layers with a 3$\times$3 convolution and 64 kernels, split in five blocks with different dilation rates, ranging from 1 to 8. The feature maps at the end of each block of each pathway were concatenated in the third dimension, resulting in a feature map with 640 kernels, and were passed to two convolutional layers with a 1$\times$1 convolution with 128 and 2 kernels, respectively. This resulted in a receptive field of 123$\times$123 pixels. The individual pathways were inspired by the P-net architecture.\textsuperscript{32} Fig. 2 gives an overview of the network architecture.\par
\begin{figure}
	\begin{center}
		\begin{tabular}{c} 
			\includegraphics[width=8.5cm]{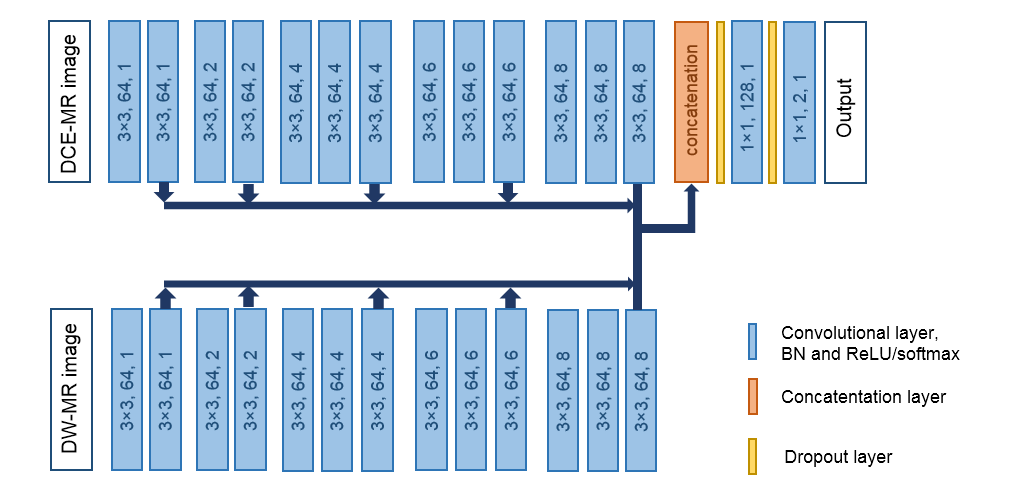}
		\end{tabular}
	\end{center}
	\caption[example]
	{ \label{fig:Fig2} 
		 Network architecture for liver metastases detection. The blue blocks represent the convolutional layers, batch normalization (BN) and ReLU or softmax activation. The size of the kernel of each convolutional layer is given in the block, followed by the number of kernels and the dilation rate of the kernel. }
		 
\end{figure}
In addition, the network was also trained and evaluated as a single pathway FCN, once with only the DCE-MR images as input images with six channels, and once with the DCE-MR and DW-MR images concatenated as input images with nine channels. In this manner the additional value of the DW-MR images and the addition of a second pathway in the detection method can be determined. The single pathway FCN was identical to the top pathway in Fig. 2. In the concatenation layer, the feature maps at the end of each block were concatenated in the third dimension.\par
Categorical cross-entropy was used as the loss function. ReLU activation and batch normalization were used in all the convolutional layers, except for the final layer which had a softmax activation. Two dropout layers were applied before and after the second-to-last layer. The dropout rate was set to 0.2. The classes were weighted based on class frequencies. He uniform\textsuperscript{33} was used as initializer and Adam as optimizer with a learning rate of 0.0001. The network was trained for 10,000 iterations, with 4 images per mini batch. The slices to train on were limited to the slices containing lesions for a more balanced data set, resulting in a total number of 1619 2D slices for training and 180 slices for validation.\par
Twenty-five patches of 128$\times$128 pixels were taken from each slice for data augmentation. The patches originated from the liver region and have overlapping areas. Data augmentation was applied by random rotation of the patches, with rotation angles of $\pm45$ degrees.\par
The data of the test set was processed by the network per 2D image slice. The input image slice consisted either of six channels of the DCE-MR phase images, three channels of DWI images, or nine channels, depending on the network. The probability output was masked by the liver segmentation, which was dilated with a 5$\times$5 structuring element. The dilation of the liver segmentation is a safety measure to ensure that small failures in the liver segmentation would not lead to undetected liver metastases.\par
For the evaluation of the detection, the masked probability output was post-processed to a binary image. All pixels with a probability output higher than the threshold of 0.5 were labelled as metastasis. Morphological closing with a structuring element of 3$\times$3$\times$3 was applied to fill any holes in the binary image. The morphological closing was followed by a morphological opening with a plus-shaped structuring element of 3$\times$3, to remove any remaining noise in the detection results. The resulting binary image was divided into separate objects representing individual metastases, using voxel clustering with 26-neighbourhood connection.

\section*{Experiments}
\subsection*{Liver segmentation}
The performance of the liver segmentation was evaluated based on the Dice similarity coefficient (DSC), the relative volume difference (RVD), and the modified Hausdorff distance (HD) at the 95\textsuperscript{th} percentile. The first annotations of the first observer of the test set (O1.1) are used to evaluate the automatic segmentation, since these were made by the same observer in the same session as the annotations of the training and validation sets.\par
DSC:  $(2*X\cap Y)/(X+Y)$, where $X$ is the automatic segmentation and $Y$ annotation O1.1.\par
RVD:$(X-Y)/Y*100\% $ \par
HD at 95\textsuperscript{th} percentile: $ max(h_{95}(X,Y),h_{95}(Y,X))$ , with $h_{95}(X,Y)= K_{(x \in X)} ^{95^{th}} min_{(y\in Y)}\|y-x\|$. Here, $X$ is the set of boundary points \{x1, .. xN\} of the automatic segmentation result, and Y is the set of boundary points \{y1, .. yN\} of annotation O1.1. K\textsuperscript{95\textsuperscript{th}}  denotes the 95\textsuperscript{th} ranked value in the set of distances between all boundary points in $X$ and the closest boundary points in $Y$.\textsuperscript{34}\par 
These three metrics were computed on the predicted segmentation results. In addition, the three metrics were computed for the second annotations of the first observer (O1.2) and the annotations of the second observer (O2) to obtain the intra- and inter-observer agreement, respectively.

\subsection*{Liver metastases detection}

A liver metastasis was considered detected, and thus a true positive object, when the manual annotation and the predicted segmentations had an overlap greater than 0. \par
For the evaluation of the liver metastases detection, the true positive rate (TPR) and the number of false positives per case (FPC) were reported. The TPR was calculated as the number of true positive objects divided by the total number of true lesion objects. The FPC was calculated as the number of objects not overlapping with any true metastasis object. In addition, the TPR and FPC were given for several thresholds in a Free-response Receiver Operating Characteristic (FROC) curve.\par
The same metrics were applied to the segmentation results of the single pathway networks.\par
An expert radiologist verified the results and determined the underlying physiology of selected false positive objects.

\section*{Results}
Fig. 3 shows some examples of the late arterial phase of the DCE-MRI, the DW-MRI with b-value 150 s/mm\textsuperscript{2}, the manually annotated liver and metastases, and the automatic segmentation of the liver with the detection of the liver metastases involving both DCE-MR and DW-MR images in the dual pathway FCN. It shows good liver and lesion segmentations, with a false positive object in the last row, which is verified to be a cyst.
\begin{figure}
	\begin{center}
		\begin{tabular}{c} 
			\includegraphics[width=8.5cm]{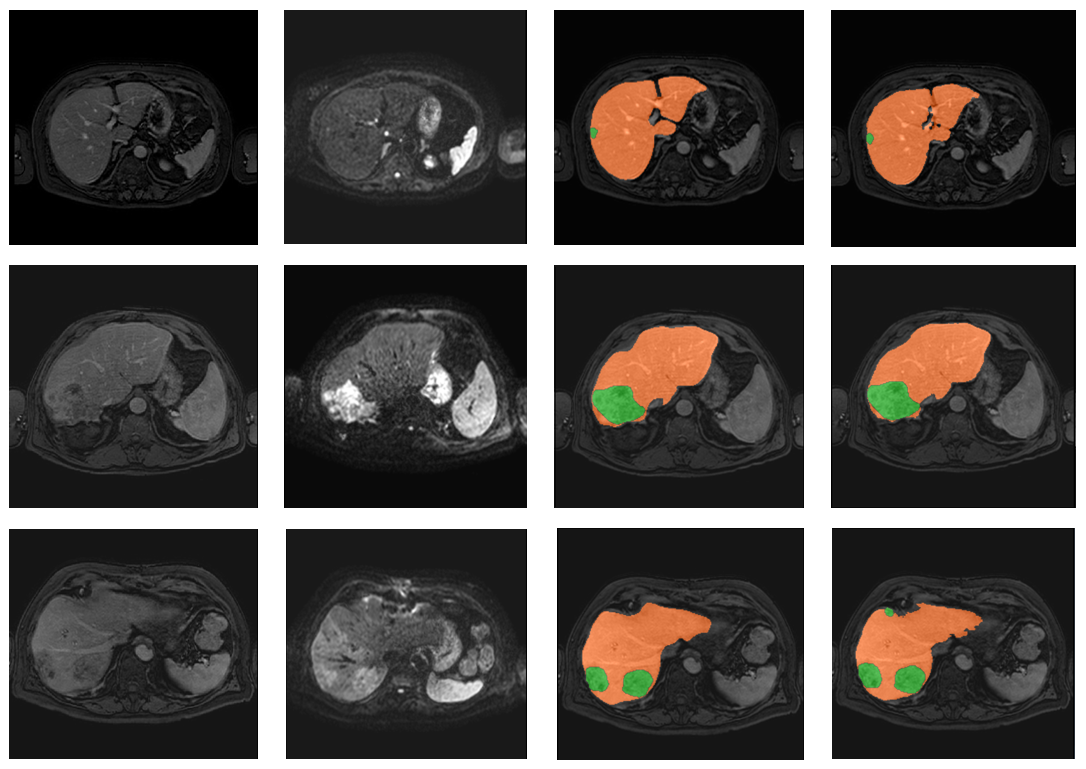}
		\end{tabular}
	\end{center}
	\caption[example]
	{ \label{fig:Fig3} 
		 Three examples of liver and metastases segmentations. From left to right each column represents a late arterial phase DCE-MR image, the registered DW-MR image with b-value 150 s/mm², the manually annotated liver (orange) and metastases (green), and the automatically segmented liver (orange) and metastases (green). The bottom row shows a false positive object in the anterior side of the liver for the automatic detection, which is a cyst.}
		 
\end{figure}
\begin{figure}
	\begin{center}
		\begin{tabular}{c} 
			\includegraphics[width=8.5cm]{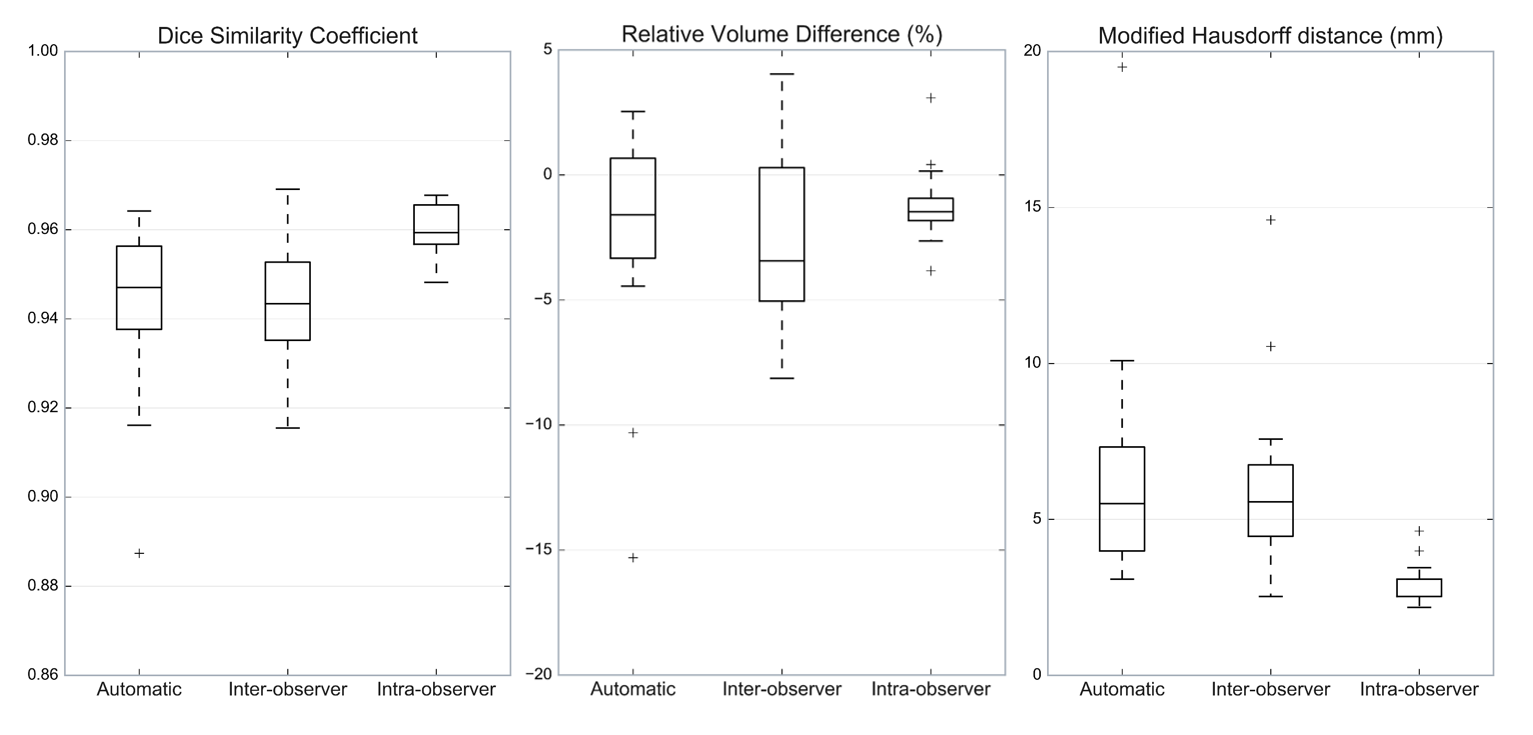}
		\end{tabular}
	\end{center}
	\caption[example]
	{ \label{fig:Fig4} 
		 Boxplots of the Dice similarity coefficient, relative volume difference (\%), and modified Hausdorff distance (mm).}
		 
\end{figure}
\subsection*{Liver segmentation}
The median DSC is 0.95 for the automatic segmentation, 0.94 for the inter-observer agreement, and 0.96 for the intra-observer agreement. The median RVD is -1.6\% for the automatic segmentation, -3.4\% for the inter-observer agreement, and -1.5\% for the intra-observer agreement. The median HD is 5.5 mm for the automatic segmentation, 5.6 mm for the inter-observer agreement, and 3.1 mm for the intra-observer agreement. The distributions of the DSC, RVD, and modified HD are given in boxplots in Fig. 4.\par

In the boxplots, annotations O1.1 are used to evaluate the automatic liver segmentations. Some variance is present between the different annotations of the observers and therefore we also compared the automatic method to all manual annotations. The median [interquartile range (IQR)] values of the DSC, RVD, and modified HD for the automatic segmentation, the inter-observer agreements, and the intra-observer agreement using the three manual annotations are given in Table 1. The upper three rows give the same results as the boxplots and the lower three rows give additional results. \par

Example results of the automatic liver segmentation are shown in the last column of Figure 3 and in Figure 8. \par

\begin{table}[b]
	\centering
	\caption{ Median and interquartile range for the Dice similarity coefficient (DSC), relative volume difference (RVD), and modified Hausdorff distance (HD) are given. In each row of the first column, the latter observer is used as the reference. The upper three rows are the same results as given in Figure 4. The lower three rows are additional results.}
	\label{Table1}
	\begin{tabular}{@{}llll@{}}
		\toprule
		     & \multicolumn{1}{c}{\textbf{DSC}} & \multicolumn{1}{c}{\textbf{RVD(\%)}} & \multicolumn{1}{c}{\textbf{HD (mm)}}  \\ \midrule
		\begin{tabular}[c]{@{}l@{}}Automatic\\ - O1.1\end{tabular}                   & 0.95 [0.93–0.96]                          & -1.6 [-3.5–0.8]                    & 5.5 [4.0–8.0]                         \\
		\begin{tabular}[c]{@{}l@{}}Inter-observer\\ O2 - O1.1\end{tabular}
                      & 0.94 [0.93–0.95]                         & -3.4 [-5.1–0.6]                          & 5.6 [4.3–6.8]                      \\
		\begin{tabular}[c]{@{}l@{}}Intra-observer\\ O1.2 - O1.1\end{tabular}
               & 0.96 [0.96–0.97]                           &-1.5 [-1.9– -0.9]                           & 3.1 [2.5–3.1]                            \\ \midrule
		\begin{tabular}[c]{@{}l@{}}Automatic\\ - O1.2\end{tabular}                        & 0.95 [0.93–0.96]                           & -0.1 [-2.8–2.2]                           & 4.9 [3.7–8.0]   \\
		\begin{tabular}[c]{@{}l@{}}Automatic\\ - O2\end{tabular}                & 0.95 [0.93–0.95]                          & 0.5 [-3.8–5.5]                           & 6.2 [4.9–9.6]                 \\ 
		\begin{tabular}[c]{@{}l@{}}Inter-observer\\ O2 - O1.2\end{tabular}
                   & 0.95 [0.94–0.96]                           & -2.5 [-4.0–1.6]                           & 4.4 [3.7–4.8]                         \\
	\bottomrule
	\end{tabular}
\end{table}

\subsection*{Liver metastases detection}
Fig. 5 shows the TPR and the FPC for all subjects. The average TPR for the single pathway network using only DCE-MRI is 0.645 and for using both DCE-MRI and DW-MRI is 0.722. The average TPR is 0.998 for the dual pathway network using both DCE-MRI and DW-MRI as input images. The median FPC is, 5 for the single pathway network using only DCE-MRI, 6 when using both MR sequences, and 2 for the dual pathway network. The single pathway has 111 and 146 false positive objects in total, using only DCE-MRI and both MR sequences, respectively. While the dual pathway has 59 false positive objects in total.\par

\begin{figure}
	\begin{center}
		\begin{tabular}{c} 
			\includegraphics[width=8.5cm]{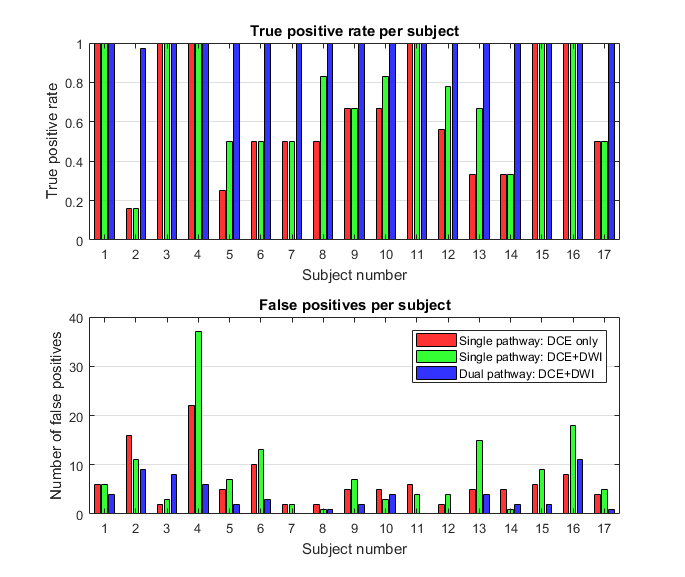}
		\end{tabular}
	\end{center}
	\caption[example]
	{ \label{fig:Fig5} 
		 The true positive rate and the number of false positives per subject for the single pathway and the dual pathway networks.}
		 
\end{figure}

Fig. 6 shows the number of detected metastases relative to the total number of metastases per size category. Both the single pathway networks fail on the smaller metastases in particular. \par

\begin{figure}
	\begin{center}
		\begin{tabular}{c} 
			\includegraphics[width=8.5cm]{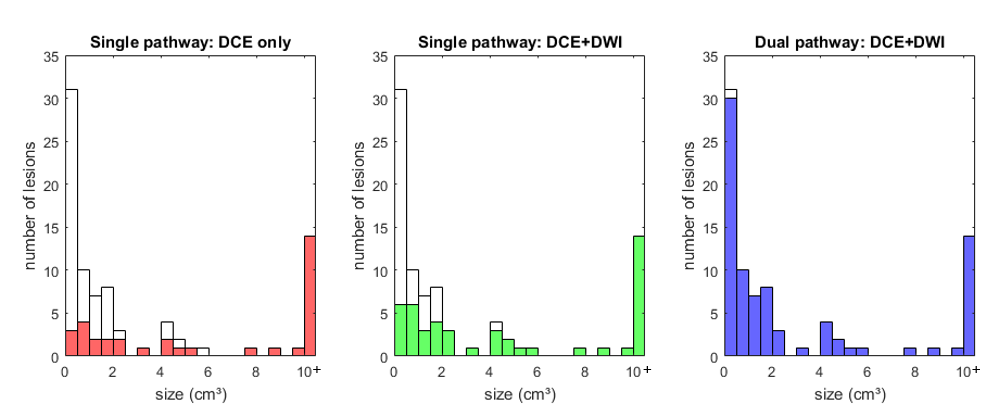}
		\end{tabular}
	\end{center}
	\caption[example]
	{ \label{fig:Fig6} 
		 The number of lesions plotted against the size of the lesions. The colored part represents the detected lesions and the white part the missed lesions at a threshold of 0.50.}
		 
\end{figure}

Fig. 7 shows the FROC curve of the mean TPR versus the median FPC for different thresholds T applied to the output of both networks. The thresholds range from 0.90 to 0.00 with steps of 0.10. At the highest threshold (T=0.90) only 42\% of the metastases was detected with a median FPC of one (in total, 31 false positive objects are present) using only the DCE-MR images, and 65\% of the metastases was detected with a median FPC of one (in total, 25 false positive object are present) using both MR sequences in the single pathway. For the dual pathway, 96\% of the metastases are detected, with a median FPC of one (in total, 19 false positive objects are present). . At the lowest threshold (T=0.0) 84\% is detected with only DCE-MR images, 91\% is detected with both MR sequences in the single pathway, and all the lesions are detected with the dual pathway network, but all at the cost of many false positives. \par

\begin{figure}
	\begin{center}
		\begin{tabular}{c} 
			\includegraphics[width=8.5cm]{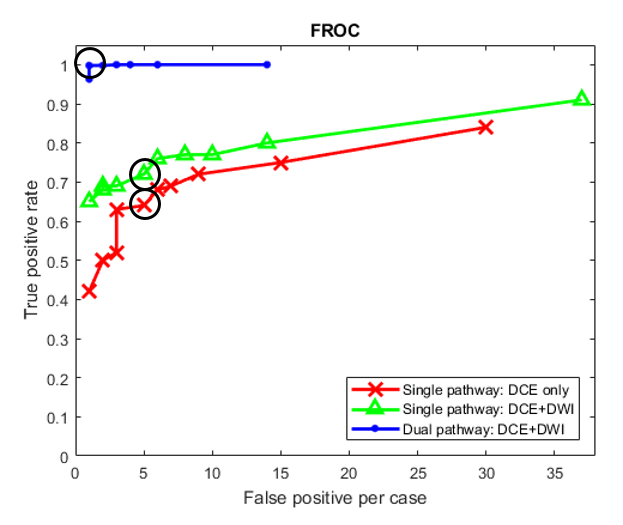}
		\end{tabular}
	\end{center}
	\caption[example]
	{ \label{fig:Fig7} 
		  FROC curve of the mean true positive rate and the median number of false positives per case for threshold T ranging from 0.90 to 0.0 in steps of 0.10. The circles represent the TPR and FPC at threshold T=0.50.}
		 
\end{figure}

Fig. 8 shows five examples of the lesion detection results including the liver segmentation result. The green pixels are true positive, blue pixels are false positive, and red pixels are false negative. Note that true positive objects are connected components that overlap with a manually annotated metastasis. A true positive object can nonetheless have some undersegmentation (red pixels, example indicated with arrow head) or oversegmentation (blue pixels, example indicated with arrow).

\begin{figure}
	\begin{center}
		\begin{tabular}{c} 
			\includegraphics[width=8.5cm]{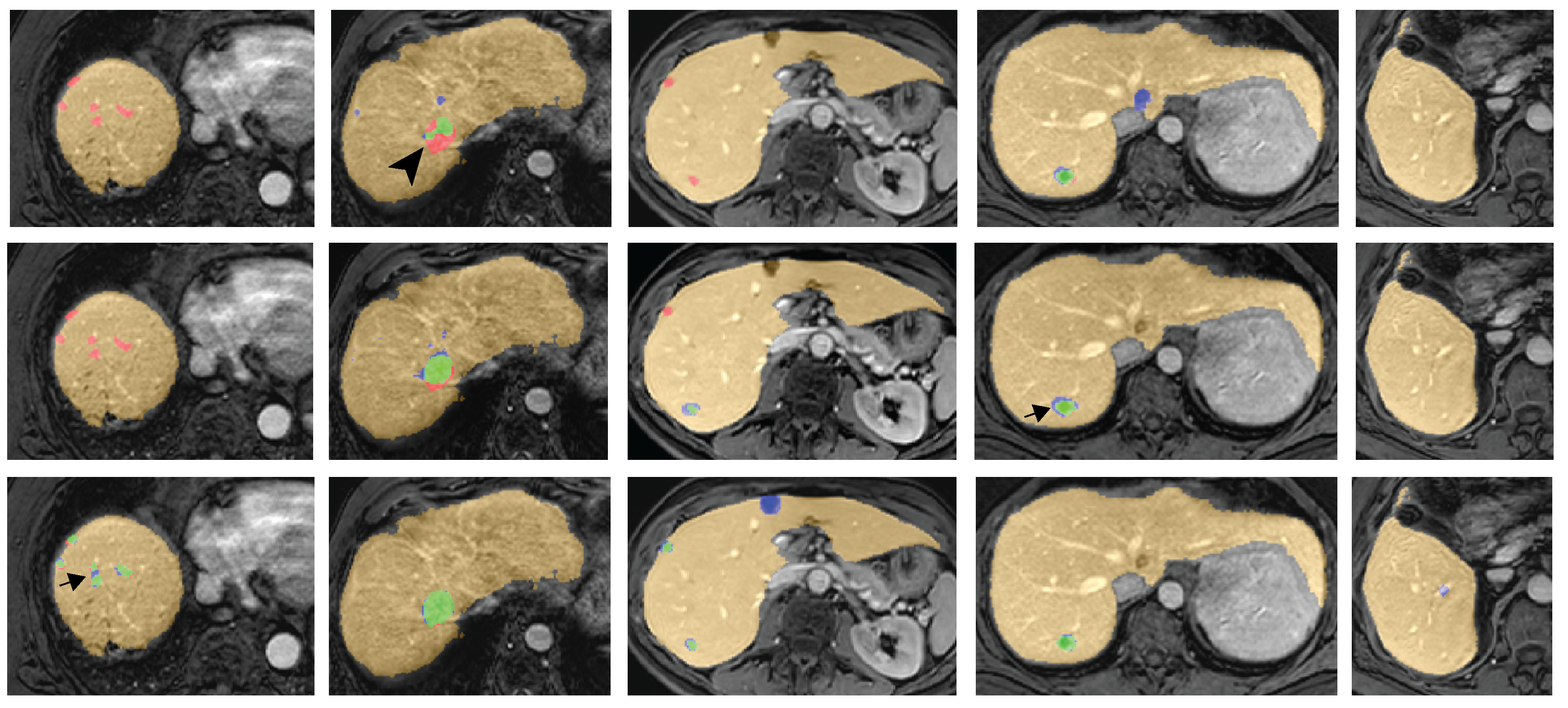}
		\end{tabular}
	\end{center}
	\caption[example]
	{ \label{fig:Fig8} 
		  Five examples of the lesion detection results. The first row shows results of the single pathway using only DCE-MRI, the second row the results of the single pathway using DCE-MRI and DW-MRI, and the third row shows the results of the dual pathway network. Green pixels are true positive, red are false negative, blue are false positive pixels. The orange tissue represents the automatic liver segmentation. The arrows indicate oversegmentation and the arrow head indicates undersegmentation. From left to right the subjects 2, 4, 5, 10, and 17 are shown.}
		 
\end{figure}

\section*{Discussion}
\subsection*{Liver segmentation}
Our proposed liver segmentation method is able to provide high quality liver segmentations using DCE-MR images. The method is able to deal with common difficulties in MR segmentation as well as the presence of lesions.\par
The automatic method performs well on the test set of annotation O1.1. The results of the automatic method are similar to the inter-observer results, as can be seen in the top rows of Table 1. Observer 1 annotated the training set so it is not surprising that the automatic segmentations do well on the test set, which was annotated by the same observer, in the same annotation session. There is some intra-observer variance between O1.1 and O1.2 and the second annotations appear systematically smaller than the first annotations, since the RVD values involving O1.1 are mostly negative. Still, the automatic method also performs well when evaluated with annotations O1.2, as can be seen in the bottom rows of Table 1. Annotations O2 were used to calculate the inter-observer variance, and were additionally used to evaluate the automatic method with another observer to verify whether the method generalizes well. The results using O2 for evaluation are still good, although slightly worse than the automatic method evaluated with O1. This is as expected, because the first observer annotated the training and validation sets.\par
The automatic liver segmentation encountered few problems. However, an outlier in all three metrics was found in one case, where a large lesion (337 ml) occurred at the edge of the liver. This outlier has an RVD of -15\%, as a result of undersegmentation of the liver in this lesion area. Even though the lesion is bigger than the receptive field of the neural network, the method accomplished to recognize half of the lesion tissue as part of the liver. However, large lesions at the edge of the liver are a potential pitfall for the proposed liver segmentation method.\par
It is hard to compare the liver segmentation results with other methods, since the data sets and MRI sequences used are not the same. A rough comparison with recently proposed automatic liver MR segmentation algorithms\textsuperscript{17,18,25} can be made by considering the reported values for the DSC. For the mentioned studies, the DSC ranged from 0.87 to 0.91, while we report a DSC of 0.95.

\subsection*{Liver metastases detection}
The detection of all liver metastases is of great importance for adequate treatment.35 The detection method with only DCE-MR images has an average TPR of 0.645 and a median of 5 false positives per case. The detection method with the single pathway using both MR sequences is more effective, with an average TPR of 0.722 and a median of 6 false positives per case. The dual pathway detection method is able to detect almost all metastases at a threshold level of 0.50 with a median of 2 false positives per case. This shows the importance of adding DW-MR images to the lesion detection method and processing the two MR sequences in separate pathways. DW-MR images are highly sensitive for lesions and in combination with the DCE-MR images the method is able to detect metastases at a high detection rate and a low number of false positives. Fig. 8 shows some visual examples of the differences in detection of these three networks.\par
The usage of the two MR sequences requires the registration of the DW-MR images to the DCE-MR image space. Ten registrations failed, which led to the exclusion of these images from the study. This is likely caused by the free-breathing protocol used for the DW-MR images that may result in deformations that cannot be corrected by the registration method. A more dedicated registration approach or a breath-hold DW-MR acquisition may improve the robustness of the method.\par
One small metastasis is missed by the dual pathway network. The metastasis is only 5 pixels in size and was detected by the dual pathway with a probability higher than 0.50. However, it is deleted during the binary opening in the post-processing step, where it is seen as noise. This reduction of noisy false positives comes at the cost of missing very small metastases.\par
The missed metastases by the single pathway networks are mostly smaller than 2 cm³, as can be seen in Fig. 6. These smaller metastases may differ in appearance from larger ones, because certain features, such as rim enhancement, cannot be expressed in only a few voxels of the DCE-MR images. The network with only DCE-MR input images is uncertain about these metastases, which leads to low outcome probabilities. However, the single pathway with both DCE-MR and DWI-MR is also less able to detect these smaller metastases. The network seems to fail to create adequate feature maps to detect small metastases, when it combines the two MR sequences in the first convolutional layer. The dual pathway network is allowed to create feature maps for the two MR sequences separately. The phases of the DCE-MR images are correlated over time, as are the three b-value images of the DW-MR. Both MR sequences contain useful characteristics.  In the dual pathway network, the feature maps are solely composed of the MR sequence presented to that pathway, representing those characteristics. And the feature maps are combined at the end. This might explain the difference the detection of the small metastases of the two networks. In addition, the dual pathway has two times more parameters as the single pathway.  Fig. 8, first column, shows some examples of sub-centimeter sized metastases.\par
The dual pathway network detects almost all metastases, but also incorrectly marks other objects as metastases. Most of these false positives are caused by objects which are not or scarcely represented in the training set and have thus an appearance unknown to the network. These unknown appearances could be other lesions and liver conditions, such as cirrhosis or an inhomogeneous fat distribution in the liver. The network has mostly seen healthy liver parenchyma with metastases during training and has learned to distinguish these, but is uncertain about conditions and lesions not seen during training. Fig. 3 (last row) and Fig. 8 (third column) show examples of cysts marked as a metastasis by the detection method.\par
Inspection of the 19 false positive objects for a threshold level of 0.90 in the post-processing step reveals that 10 of the false positives are lesions: 9 cysts and 1 hemangioma. Three of the false positives are blood vessels and one false positive is because of an inhomogeneous fat distribution in the liver. Fig. 8 (last column) shows an example of a blood vessel marked as a metastasis. Three other false positives are located on the edge of the liver, which sometimes has a higher intensity than the rest of the liver (e.g. Fig. 3 second column). Furthermore, motion artifacts can be the cause of a false positive object, which is the case for two false positives.\par
The morphological operations of on the detection results were carefully designed for the tasks described. However, post hoc analysis revealed that the liver mask dilation with a 5×5 structuring element did not seem to be necessary for this data set, since all liver metastases were included in the original liver mask. On the other hand, the morphological closing and opening of the lesion detection results did have an impact on the results of the dual pathway method. Without these morphological operations the median FPC would increase to 8 instead of 2. Nevertheless, the morphological operations removed one small lesion, from a total of 86, which was otherwise detected.\par
These CNN models are trained on images from a specific MR protocol from one clinic. Like other CNN models, there might be a drop in performance when the model is used on data from another scanner or another clinic\textsuperscript{36}. To avoid this problem, the CNN should be fine-tuned on data similar to the test data, e.g. by using transfer learning. \par
This work could also be expanded to 3D input images, adding more spatial information, which might improve the results. However, this would require more training data and computational power to train and test a network with more parameters. 

\section*{Conclusions}
An automatic liver segmentation with a similarity index comparable to that of the inter-observer agreement, is obtained from dynamic contrast enhanced MR images with a fully convolutional neural network. The method can accurately segment livers irrespective of the liver condition or the presence of lesions. Only in the event of lesions larger than the receptive field of the fully convolutional neural network, parts of the liver are missed.\par
The proposed dual pathway metastases detection method, based on dynamic contrast enhanced MR and diffusion weighted MR images, successfully detects 99.8\% of the liver metastases at the cost of a median of two false positives per case. This will aid radiologists to locate metastases quickly.

\begin{acknowledgements}
The authors thank Dr. Ashis Kumar Dhara and Prof. Dr. Robin Strand from the Centre of Image Analysis of Uppsala University for discussions regarding the lesion detection method.\par
This work was financially supported by the project IMPACT (Intelligence based iMprovement of Personalized treatment And Clinical workflow supporT) in the framework of the EU research programme ITEA3 (Information Technology for European Advancement).

\end{acknowledgements}

\begin{copyright}
Copyright 2019 Society of Photo-Optical Instrumentation Engineers. One print or electronic copy may be made for personal use only. Systematic reproduction and distribution, duplication of any material in this paper for a fee or for commercial purposes, or modification of the content of the paper are prohibited.
\end{copyright}

\section*{References}
\small{
1.		T. J. Vogl, N. N. N. Naguib, S. Zangos, K. Eichler, A. Hedayati, and N. E. A. Nour-Eldin, “Liver metastases of neuroendocrine carcinomas: Interventional treatment via transarterial embolization, chemoembolization and thermal ablation,” \textit{Eur. J. Radiol.}, 72(3), pp. 517–528, 2009.\par
2.		E. Van Cutsem, B. Nordlinger, R. Adam, C. H. Köhne, C. Pozzo, G. Poston, M. Ychou, and P. Rougier, “Towards a pan-European consensus on the treatment of patients with colorectal liver metastases,” \textit{Eur. J. Cancer}, 42(14), pp. 2212–2221, 2006.\par
3.	P. J. Robinson, “The early detection of liver metastases,” \textit{Cancer Imaging}, 2, pp. 1–3, 2002.\par
4.		Z. Bakhtiary, A. A. Saei, M. J. Hajipour, M. Raoufi, O. Vermesh, and M. Mahmoudi, “Targeted superparamagnetic iron oxide nanoparticles for early detection of cancer: Possibilities and challenges,” \textit{Nanomedicine Nanotechnology}, Biol. Med., 12(2), pp. 287–307, 2016.\par
5.		World Health Organization, “WHO Fact sheet: Cancer,” 2018.\par
6.		A. C. Silva, J. M. Evans, A. E. McCullough, M. A. Jatoi, H. E. Vargas, and A. K. Hara, “MR imaging of hypervascular liver masses: a review of current techniques.” \textit{Radiographics}, 29(2), pp. 385–402, 2009.\par
7.	J. Böttcher, A. Hansch, A. Pfeil, P. Schmidt, A. Malich, A. Schneeweiss, M. H. Maurer, F. Streitparth, U. K. Teichgräber, and D. M. Renz, “Detection and classification of different liver lesions: Comparison of Gd-EOB-DTPA-enhanced MRI versus multiphasic spiral CT in a clinical single centre investigation,” \textit{Eur. J. Radiol.}, 82(11), pp. 1860–1869, 2013.\par
8.		V. Vilgrain, M. Esvan, M. Ronot, A. Caumont-prim, C. Aubé, and G. Chatellier, “A meta-analysis of diffusion-weighted and gadoxetic acid-enhanced MR imaging for the detection of liver metastases,” \textit{Eur. Radiol.}, 26, pp. 4595–4615, 2016.\par
9.		K. Holzapfel, M. J. Eiber, A. A. Fingerle, M. Bruegel, E. J. Rummeny, and J. Gaa, “Detection , classification , and characterization of focal liver lesions: Value of diffusion-weighted MR imaging , gadoxetic acid-enhanced MR imaging and the combination of both methods,” \textit{Abdom. Imaging}, 37, pp. 74–82, 2012.\par
10.		C. Kenis, F. Deckers, B. De Foer, F. van Mieghem, S. Van Laere, and M. Pouillon, “Diagnosis of liver metastases: Can diffusion-weighted imaging be used as a stand alone sequence?,” \textit{Eur. J. Radiol.}, 81, pp. 1016–1023, 2012.\par
11.		K. M. Elsayes, V. R. Narra, Y. Yin, G. Mukundan, M. Lammle, and J. J. Brown, “Focal hepatic lesions: Diagnostic value of enhancement pattern approach with contrast enhanced 3D gradient-echo MR imaging,” \textit{Radiographics}, 25(5), pp. 1299–1320, 2005.\par
12.	C. Schmid-Tannwald, S. Thomas, M. K. Ivancevic, F. Dahi, C. Rist, I. Sethi, and A. Oto, “Diffusion-weighted MRI of metastatic liver lesions: is there a difference between hypervascular and hypovascular metastases?,” \textit{Acta radiol.}, 55(5), pp. 515–523, 2014.\par
13.		G. Litjens, T. Kooi, B. E. Bejnordi, A. A. A. Setio, F. Ciompi, M. Ghafoorian, J. A. W. M. Van Der Laak, B. Van Ginneken, and C. I. Sánchez, “A survey on deep learning in medical image analysis,” Med. Image Anal., 42, pp. 60–88, 2017.\par
14.	P. F. Christ, “Liver Tumor Segmentation Challenge,” 2017. [Online]. Available: lits-challenge.com.\par
15.		F. Lopez-Mir, V. Naranjo, J. Angulo, M. Alcañiz, and L. Luna, “Liver segmentation in MRI: A fully automatic method based on stochastic partitions,” \textit{Comput. Methods Programs Biomed.}, 114(1), pp. 11–28, 2014.\par
16.	H. Masoumi, A. Behrad, M. A. Pourmina, and A. Roosta, “Automatic liver segmentation in MRI images using an iterative watershed algorithm and artificial neural network,” \textit{Biomed. Signal Process. Control}, 7(5), pp. 429–437, 2012.\par
17.		H. T. Huynh, N. Le-Trong, P. T. Bao, A. Oto, and K. Suzuki, “Fully automated MR liver volumetry using watershed segmentation coupled with active contouring,” \textit{Int. J. Comput. Assist. Radiol. Surg.}, 12(2), pp. 235–243, 2017.\par
18.		E. Dura, J. Domingo, E. Göçeri, and L. Martí, “A method for liver segmentation in perfusion MR images using probabilistic atlases and viscous reconstruction,” \textit{Pattern Anal. Appl.}, 21(4), pp. 1083–1095, 2018.\par
19.	G. Chartrand, T. Cresson, R. Chav, A. Gotra, A. Tang, and J. A. De Guise, “Liver Segmentation on CT and MR Using Laplacian Mesh Optimization,” \textit{IEEE Trans. Biomed. Eng.}, 64(9), pp. 2110–2121, 2017.\par
20.		A. Ben-Cohen, E. Klang, A. Kerpel, E. Konen, M. M. Amitai, and H. Greenspan, “Fully convolutional network and sparsity-based dictionary learning for liver lesion detection in CT examinations,” \textit{Neurocomputing}, 275, pp. 1585–1594, 2018.\par
21.	P. F. Christ, M. E. A. Elshaer, F. Ettlinger, S. Tatvarty, M. Bickel, P. Bilic, M. Rempfler, M. Armbruster, F. Hofmann, M. D’Anastasi, W. H. Sommer, S.-A. Ahmadi, and B. H. Menze, “Automatic liver and lesion segmentation in CT using cascaded fully convolutional neural networks and 3D conditional random fields,” in \textit{Lecture Notes in Computer Science MICCAI 2016}, 9901(2), pp. 415–423, 2016.\par
22.	M. Bilello, S. B. Gokturk, T. Desser, S. Napel, R. B. Jeffrey, and C. F. Beaulieu, “Automatic detection and classification of hypodense hepatic lesions on contrast-enhanced venous-phase CT,” \textit{Med. Phys.}, 31(9), pp. 2584–2593, 2004.\par
23.		E. Vorontsov, A. Tang, C. Pal, and S. Kadoury, “Liver lesion segmentation informed by joint liver segmentation,” in \textit{Proceedings of IEEE 15th International Symposium on Biomedical Imaging (ISBI 2018)}, pp. 1332–1335, 2018.\par
24.		A. L. M. Pavan, M. Benabdallah, M.-A. Lebre, D. Rodigrues de Pina, F. Jaziri, A. Vacavant, A. Mtibaa, H. Mohamed Ali, M. Grand-Brochier, H. Rositi, B. Magnin, A. Abergel, and P. Chabrot, “A Parallel Framework for HCC Detection in DCE-MRI Sequences with Wavelet-Based Description and SVM classification,” in\textit{ Proceedings of the 33rd Annual ACM Symposium on Applied Computing}, pp. 14–21, 2018.\par
25.		P. F. Christ, F. Ettlinger, F. Grün, M. E. A. Elshaer, J. Lipková, S. Schlecht, F. Ahmaddy, S. Tatavarty, M. Bickel, P. Bilic, M. Rempfler, F. Hofmann, M. D’Anastasi, S.-A. Ahmadi, G. Kaissis, J. Holch, W. H. Sommer, R. Braren, V. Heinemann, and B. H. Menze, “Automatic Liver and Tumor Segmentation of CT and MRI Volumes Using Cascaded Fully Convolutional Neural Networks,” \textit{arXiv Prepr. arXiv1702.05970}, pp. 1–20, 2017.\par
26.	M.J.A. 	Jansen, H.J. Kuijf, J.P.W Pluim, “Optimal input configuration of dynamic contrast enhanced MRI in convolutional neural networks for liver segmentation,” in \textit{Proceedings of SPIE 10949, Medical Imaging 2019: Image Processing.} 109491V, 2019\par
27.		M.J.A. Jansen, H.J. Kuijf, W.B. Veldhuis, F.J. Wessels, M.S. van Leeuwen, and J.P.W. Pluim, “Evaluation of motion correction for clinical dynamic contrast enhanced MRI of the liver,” \textit{Phys. Med. Biol.}, 62(19), pp. 7556–7568, 2017.\par
28.		S. Klein, M. Staring, K. Murphy, M. A. Viergever, and J. P. W. Pluim, “elastix: A toolbox for intensity-based medical image registration,” \textit{IEEE Trans. Med. Imaging}, 29(1), pp. 196–205, 2010.\par
29.		F. Yu and V. Koltun, “Multi-Scale Context Aggregation by Dilated Convolutions,” in \textit{ICLR}, 2016.\par
30.		F. Milletari, N. Navab, and S.-A. Ahmadi, “V-Net: Fully Convolutional Neural Networks for Volumetric Medical Image Segmentation,” in \textit{Fourth International Conference on 3D Vision (3DV)}, pp. 1–11, 2016.\par
31.		X. Glorot and Y. Bengio, “Understanding the difficulty of training deep feedforward neural networks,” in \textit{Proceedings of the 13th International Conference on Artificial Intelligence and Statistics (AISTATS) 2010}, 9, pp. 249–256, 2010.\par
32.	G. Wang, W. Li, M. A. Zuluaga, R. Pratt, P. A. Patel, M. Aertsen, T. Doel, L. Anna, J. Deprest, and T. Vercauteren, “Interactive medical image segmentation using deep learning with image-specific fine-tuning,” \textit{IEEE Trans. Med. Imaging}, 37(7), pp. 1562–1573, 2018.\par
33.		K. He, X. Zhang, S. Ren, and J. Sun, “Delving deep into rectifiers: Surpassing human-level performance on imagenet classification,” in \textit{Proceedings of the IEEE International Conference on Computer Vision}, pp. 1026–1034, 2015.\par
34.	D. P. Huttenlocher, G. A. Klanderman, and W. J. Rucklidge, “Comparing images using the Hausdorff distance,” \textit{IEEE Trans. Pattern Anal. Mach. Intell.}, 15(9), pp. 850–863, 1993.\par
35.		G. Masi, L. Fornaro, C. Caparello, and A. Falcone, “Liver metastases from colorectal cancer: how to best complement medical treatment with surgical approaches,” \textit{Future Oncol.}, 7(11), 2011.\par
36.	 Kuijf, H. J. et al. “Standardized Assessment of Automatic Segmentation of White Matter Hyperintensities; Results of the WMH Segmentation Challenge,” \textit{IEEE Trans. Med. Imaging}, Early Access, 2019\par
}

\end{document}